# Label Tree Embeddings for Acoustic Scene Classification

Huy Phan
Institute for Signal Processing
University of Lübeck
phan@isip.uni-luebeck.de

Lars Hertel
Institute for Signal Processing
University of Lübeck
hertel@isip.uni-luebeck.de

Marco Maass
Institute for Signal Processing
University of Lübeck
maass@isip.uni-luebeck.de

Philipp Koch
Institute for Signal Processing
University of Lübeck
koch@isip.uni-luebeck.de

Alfred Mertins
Institute for Signal Processing
University of Lübeck
mertins@isip.uni-luebeck.de

## ABSTRACT

We present in this paper an efficient approach for acoustic scene classification by exploring the structure of class labels. Given a set of class labels, a category taxonomy is automatically learned by collectively optimizing a clustering of the labels into multiple meta-classes in a tree structure. An acoustic scene instance is then embedded into a low-dimensional feature representation which consists of the likelihoods that it belongs to the meta-classes. We demonstrate state-of-the-art results on two different datasets for the acoustic scene classification task, including the DCASE 2013 and LITIS Rouen datasets.

## Keywords

acoustic scene classification; label tree embedding; spectral clustering

## 1. INTRODUCTION

Acoustic scene classification (ASC) is an important problem of computational auditory scene analysis [26, 16]. Solving this problem will allow a device to recognize a surrounding environment via the sound it captures, and hence, enables a wide range of applications, such as surveillance [22], robotic navigation [8], and context-aware services [27, 11]. A recognized scene can also be used as a prior information to improve the performance of sound event detection [13].

Excluding the background noise, an acoustic scene usually involves various kinds of foreground sounds. Due to its complex sound composition, it is challenging to obtain a good representation for classification. Different features adapted from the related problems, such as speech recognition and audio event classification, have been used to characterize an acoustic scene, for instance MFCC [19, 24] and Gammatone filters [25]. Some hand-crafted features tailored for the task have also been proposed and demonstrated good performance, like Histogram of Oriented Gradients (HOG) [23, 4, 28] and Gabor dictionaries [17]. At a higher semantic level, foreground sound events [2, 14, 7], background noise [9], and their combination [28] can be used as a footprint to represent a scene [2, 14, 7].

However, most (if not all) previous methods considered the "flat" classification scheme. Thus far, to the best of the authors' knowledge, no studies have explored the structured nature of the scene categories for classification. In this work, we incorporate a class taxonomy by learning to group similar categories into meta-classes on a tree structure. Going beyond that, we construct explicit embeddings to map each acoustic scene instance into the semantic space that underlies the class hierarchy. It turns out that two similar scene instances are expected to be close to each other in the semantic space. We study the class hierarchy learned from the acoustic scene data themselves as well as the one learned from external speech data [20, 21]. Both of them show good empirical performance even with simple linear classifiers. In addition, combining them with a simple fusion scheme leads to state-of-the-art performance on both target datasets: DCASE 2013 [24] and LITIS Rouen datasets [23].

## 2. THE PROPOSED APPROACH

In this section, we firstly present the framework to learn the label trees and the label tree embeddings for feature mapping. Afterwards, we elaborate different label tree embeddings derived from the framework and the final classification step.

### 2.1 Learning a Label Tree

Consider a database (e.g. scene database) with the label set $\mathcal{L} = \{1, \ldots, C\}$ where $C$ indicates the number of target categories. In order to explore the structure of class labels, we learn a label tree similar to [3]. The learning algorithm collectively partitions the label set into disjoint subsets in such a way that they are easy to distinguish from one another. Given the set of samples $\mathcal{S} = \{(\mathbf{x}_n, c_n)\}_{n=1}^{|\mathcal{S}|}$ extracted from the training data, where $\mathbf{x} \in \mathbb{R}^M$ denotes the vector of some $M$ low-level features, $c \in \mathcal{L}$ indicates the class label, and $|\cdot|$ represents the set cardinality.

The label tree is constructed recursively so that each node is associated with a set of class labels. Consider a node with a label set $\ell$ (and therefore, the root node is assigned with the label set $\mathcal{L}$), our goal is to split $\ell$ into two subsets $\ell^L$ and $\ell^R$ that hold the following requirements: $\ell^L \neq \emptyset$, $\ell^R \neq \emptyset$, $\ell^L \cup \ell^R = \ell$, and $\ell^L \cap \ell^R = \emptyset$. There are totally $2^{|\ell|-1}-1$ such possible partitions $\{\ell^L, \ell^R\}$. The optimal partition is then adopted such that a binary classifier designed to separate $\ell^L$ and $\ell^R$ makes as few errors as possible.

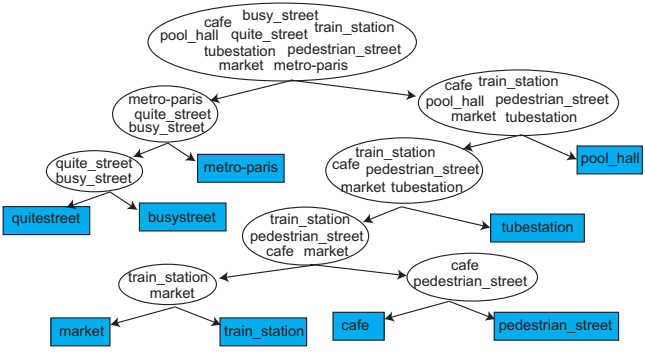

Figure 1: A subtree extracted from the label tree learned from the LITIS Rouen dataset [23].

In order to find the optimal partitioning, we rely on the multi-class confusion matrix which indicates how good a class is separated from the others. Let $\mathcal{S}^\ell \subset \mathcal{S}$ denote the set of samples corresponding to the label set $\ell$. Furthermore, suppose that we have changed and sorted the label set $\ell$ so that $\ell = \{1, \ldots, |\ell|\}$. In addition, we divide $\mathcal{S}^\ell$ into two equal halves: $\mathcal{S}^\ell_{\text{train}}$ for training a classifier and $\mathcal{S}^\ell_{\text{eval}}$ for evaluation. We train the multi-class classifier $\mathcal{M}^\ell$ using random forest classification [6] with 200 trees using $\mathcal{S}^\ell_{\text{train}}$ and then evaluate it on the evaluation set $\mathcal{S}^\ell_{\text{eval}}$ to obtain the confusion matrix $\mathbf{A} \in \mathbb{R}^{|\ell| \times |\ell|}$. Each element $\mathbf{A}_{ij}$ of the matrix $\mathbf{A}$ is computed by:

$$\mathbf{A}_{ij} = \frac{1}{|\mathcal{S}^\ell_{\text{eval},i}|} \sum_{\mathbf{x} \in \mathcal{S}^\ell_{\text{eval},i}} P(j|\mathbf{x}, \mathcal{M}^\ell). \quad (1)$$

Here, $\mathcal{S}^\ell_{\text{eval},i} \subset \mathcal{S}^\ell_{\text{eval}}$ is the set of samples with the label $i$. $P(j|\mathbf{x}, \mathcal{M}^\ell)$ denotes the probability that the classifier $\mathcal{M}^\ell$ predicts the sample $\mathbf{x}$ as class $j$. $\mathbf{A}_{ij}$ implies how likely a sample of the class $i$ is wrongly predicted to belong to the class $j$ by the classifier. Since $\mathbf{A}$ is not symmetric, we symmetrize it as

$$\bar{\mathbf{A}} = (\mathbf{A} + \mathbf{A}^\mathsf{T})/2. \quad (2)$$

Eventually, the optimal partitioning $\{\ell^L, \ell^R\}$ is selected to maximize:

$$E(\ell) = \sum_{i,j \in \ell^L} \bar{\mathbf{A}}_{ij} + \sum_{m,n \in \ell^R} \bar{\mathbf{A}}_{mn}. \quad (3)$$

By this, we tend to group the ambiguous classes into the same subset, as a result, produce two meta-classes $\{\ell^L, \ell^R\}$ that are easy to separate from each other. We apply spectral clustering [18] on the matrix $\bar{\mathbf{A}}$ to solve a relaxed version of the optimization problem in (3). The subsets $\ell^L$ and $\ell^R$ are then directed to the left and right child nodes, respectively. The splitting process is recursively repeated to grow the whole tree until a leaf node with a single class label is reached.

We demonstrate in Figure 1 a subtree extracted from the label tree learned from the LITIS Rouen dataset [23] (more details in Section 2.3).

### 2.2 Label Tree Embedding (LTE)

Via the learned label tree, we have formed $(C-1) \times 2$ meta-classes in total from the original label set $\mathcal{L}$. Two of them are associated with the left and right child nodes of one out of $(C-1)$ split nodes. For clarity, suppose that we have indexed the split nodes of the label tree as $\{\ell_i\}_{i=1}^{C-1}$. Our objective is then to learn a representation for a test sample by embedding them into the space of the meta-class labels. Formally, we then want to obtain an explicit mapping $\Psi : \mathbb{R}^M \to \mathbb{R}^{(C-1) \times 2}$ to map the test sample $\mathbf{x} \in \mathbb{R}^M$ to a feature vector $\Psi(\mathbf{x}) = \left(\psi_1^L(\mathbf{x}), \psi_1^R(\mathbf{x}), \ldots, \psi_{C-1}^L(\mathbf{x}), \psi_{C-1}^R(\mathbf{x})\right)$. The entries of $\psi_i^L(\mathbf{x})$ and $\psi_i^R(\mathbf{x})$ denote the likelihoods that $\mathbf{x}$ belongs to two meta-classes on the left and right child nodes of the split node $\ell_i$.

To obtain the likelihoods, at a split node $\ell_i$ with the optimal partition $\{\ell_i^L, \ell_i^R\}$, we train the binary random-forest classifier $\mathcal{M}_i^\ell$ with 200 trees using the whole set $\mathcal{S}^{\ell_i}$ as training data. The samples with their labels in $\ell_i^L$ are considered as negative examples and others with their labels in $\ell_i^R$ are considered as positive examples. The likelihoods are then given by:

$$\psi_i^L(\mathbf{x}) = P(\text{negative}|\mathbf{x}, \mathcal{M}^{\ell_i}), \quad (4)$$

$$\psi_i^R(\mathbf{x}) = P(\text{positive}|\mathbf{x}, \mathcal{M}^{\ell_i}). \quad (5)$$

Here, $P(\text{negative}|\mathbf{x}, \mathcal{M}^{\ell_i})$ and $P(\text{positive}|\mathbf{x}, \mathcal{M}^{\ell_i})$ are the classification probabilities outputted by $\mathcal{M}^{\ell_i}$ when evaluating on $\mathbf{x}$, thanks to the probability support of the random forest classification [6].

### 2.3 Scene and Speech LTEs

Using the above-described framework, we study following LTEs to cope with the acoustic scene classification task: (1) the LTE derived from a target scene database itself (Scene-LTE), (2) the LTE learned from an external speech data (Speech-LTE), and (3) their combination (Fusion-LTE).

**Scene-LTE.** Given a target scene database (e.g. LITIS Rouen), the Scene-LTE is formed following the framework. However, we do not consider the whole 30-second snippet of an acoustic scene instance as a sample. Instead, in order to capture meaningful events happening in a scene whose lengths are in order of hundreds of milliseconds, we use segments of length 500 ms with an overlap of 250 ms as the samples for further processing. Each segment is decomposed into 50 ms frames with 50% overlap, each of which is described by $M = 128$ Gammatone cepstral coefficients [25, 10] in the frequency range of 1-11025 Hz. A segment is then represented by a 128-dimensional feature vector computed by averaging the feature vectors of its constituent frames. Furthermore, each audio segment is labeled by the label of the scene where it is taken from.

The learned Scene-LTE is then applied on a test audio segment, resulting in a Scene-LTE feature vector. To compute the global Scene-LTE features for the 30-second scene snippet, we employ average pooling on the Scene-LTE features of its constituent segments.

**Speech-LTE.** Speech signals have been shown to bear potential to serve as a generic representation for nonspeech audio events [20, 21]. We show here that they can also been used to represent acoustic scenes. The Speech-LTE is learned from a set of *phone triplets* [21] selected from TIMIT speech data [12]. We rely on the closeness measure between a phone triplet category and a scene category for the selection. From the training 500 ms segments obtained from the target scene database, we train a 200-tree multi-class random-forest classifier $\mathcal{M}_{\text{scene}}$. The closeness $\kappa(c, y)$ of a scene category $c$ and a phone triplet category $y$ and is then computed as

$$\kappa(c,y) = \frac{1}{|\mathcal{S}^y|} \sum_{\mathbf{x}^y \in \mathcal{S}^y} P(c|\mathbf{x}^y, \mathcal{M}_{\text{scene}}). \quad (6)$$

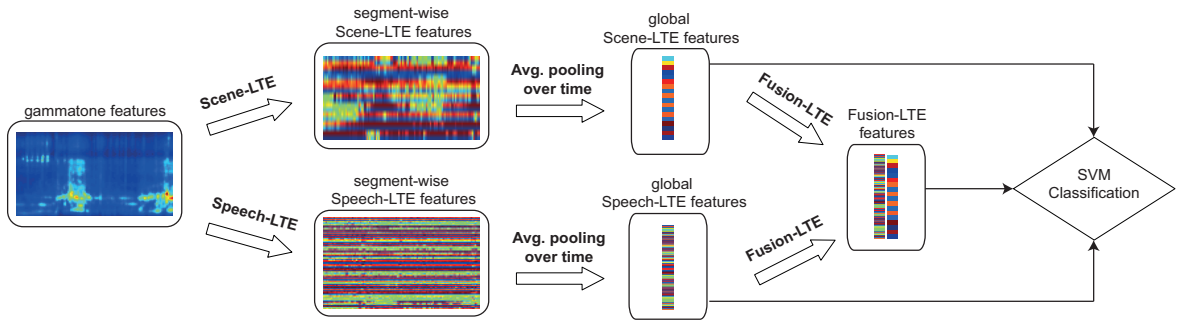

Figure 2: Overview of the proposed acoustic scene classification schemes.

Table 1: *Scene-LTE:* Performance in terms of accuracy (%) (DCASE) and F1-score (%) (LITIS).

|  | linear | $\chi^2$ | hist. | RBF |
|---|---|---|---|---|
| DCASE | 84.0±5.5 | 85.0±5.5 | 84.0±2.2 | 86.0±4.2 |
| LITIS | 94.3±0.9 | 94.9±1.0 | 94.5±1.0 | 94.8±0.8 |

Here, $\mathcal{S}^y = \{\mathbf{x}_i^y\}_{i=1}^{|\mathcal{S}^y|}$ denotes the sample set of the phone triplet category $y$. We then rank closeness measures and select top $N = \{5, 10, 15, 20, 25\}$ phone triplet categories for each scene class. The Speech-LTE is finally applied on the audio segments of a scene snippet, followed by the average pooling to produce the global Speech-LTE feature vector for the scene instance. Note that, instead of the Gammatone cepstral coefficients used in the Scene-LTE, we utilized the same low-level feature set in [21] for the Speech-LTE.

**Fusion-LTE.** In order to take advantage of representations from different perspectives, i.e. Scene-LTE and Speech-LTE, we combine them using the extended Gaussian-$\chi^2$ kernel [15] given by

$$K(\mathbf{x}_i, \mathbf{x}_j) = \exp\Big(-\sum_k \frac{1}{\bar{D}^k} D\big(\Psi^k(\mathbf{x}_i), \Psi^k(\mathbf{x}_j)\big)\Big), \quad (7)$$

where $D\big(\Psi^k(\mathbf{x}_i), \Psi^k(\mathbf{x}_j)\big)$ is the $\chi^2$ distance between the embedded scene instances $\Psi^k(\mathbf{x}_i)$ and $\Psi^k(\mathbf{x}_j)$ with respect to the $k$-th channel where $k \in \{\text{Scene-LTE}, \text{Speech-LTE}\}$. $\bar{D}^k$ is the mean $\chi^2$ distance of the embedded scene instances in training data for the $k$-th channel.

### 2.4 Final Acoustic Scene Classification

An overview of the final classification schemes is illustrated in Figure 2. For a test acoustic scene instance, we obtain its representations using the Scene-LTE and Speech-LTE as above. To extract representations for the training instances, we conducted 10-fold cross-validation on training data. Lastly, we trained the final scene classification systems using one-vs-one support vector machines (SVM) with different kernels, including linear, $\chi^2$, histogram intersection (*hist* for short), and radial basis function (RBF) kernels.

Table 2: *Speech-LTE:* Performance in terms of accuracy (%) (DCASE) and F1-score (%) (LITIS).

|  | $N$ | linear | $\chi^2$ | hist. | RBF |
|---|---|---|---|---|---|
| DCASE | 5 | 78.0±4.5 | 80.0±3.5 | 84.0±2.2 | 82.0±2.7 |
| | 10 | 84.0±2.2 | 83.0±2.7 | 86.0±6.5 | 86.0±5.5 |
| | 15 | 85.0±5.0 | 85.0±3.5 | 79.0±10.8 | 79.0±4.2 |
| | 20 | 83.0±7.6 | 85.0±6.1 | 86.0±6.5 | 83.0±7.6 |
| | 25 | 87.0±2.7 | 85.0±6.1 | 86.0±4.2 | 85.0±5.0 |
| LITIS | 5 | 85.4±1.0 | 85.9±1.3 | 86.4±0.9 | 87.9±1.4 |
| | 10 | 87.1±1.2 | 88.2±1.0 | 87.8±0.9 | 89.3±1.2 |
| | 15 | 87.8±1.0 | 88.5±1.1 | 89.1±1.2 | 89.9±1.0 |
| | 20 | 88.2±1.3 | 89.2±1.1 | 89.1±0.8 | 89.7±1.1 |
| | 25 | 88.6±1.2 | 89.7±1.1 | 89.5±0.7 | 90.1±1.2 |

Table 3: *Fusion-LTE:* Performance in terms of accuracy (%) (DCASE) and F1-score (%) (LITIS).

| $N$ | DCASE | LITIS |
|---|---|---|
| 5 | 84.0 ± 6.5 | 96.1 ± 1.0 |
| 10 | 86.0 ± 6.5 | 96.2 ± 0.9 |
| 15 | 86.0 ± 5.5 | 96.2 ± 0.9 |
| 20 | 86.0 ± 5.5 | 96.1 ± 1.0 |
| 25 | 87.0 ± 6.7 | 96.2 ± 0.9 |

For Fusion-LTE, we used nonlinear SVMs with the extended Gaussian kernel given in (7). The hyperparameters of the SVMs were tuned via 10-fold cross-validation.

## 3. EXPERIMENTS

### 3.1 Datasets

We employed the following two datasets in our experiments:

**DCASE 2013 dataset [2, 24].** This dataset was used in the DCASE 2013 challenge [24]. It consists of ten scene categories recorded in different locations in London at different time points. The dataset has two subsets: public and private subsets, each contains 100 30-second-long scene instances with ten examples for each class. The former was released during the challenge for participants to tune their classification systems. The latter was used to evaluate the submissions and also made public after the challenge. The submitted systems were evaluated with five-fold stratified cross validation on the private subset [24]. We follow the cross validation setting, however, at each time, we combined the public set and the training folds of the private set to make the training data.

**LITIS Rouen dataset [23].** This dataset contains 3026 30-second-long examples of 19 urban scene categories recorded with the total duration of 1500 minutes. Each class is specific to a location such as a train station, an airplane, or a market. To our knowledge, this is so far the largest publicly available dataset for the task. We follow the standard training/testing splits in [23] (for more details, please refer to [23]) and report average performances over 20 splits of the data.

### 3.2 Experimental Results

The classification performance obtained by Scene-LTE, Speech-LTE, and Fusion-LTE are shown in Tables 1, 2, and 3. For the DCASE dataset, the performance is reported in terms of classification accuracy as in the DCASE 2013 challenge [24] whereas we used average class-wise F1-score for the LITIS dataset since it exhibits significant imbalance in the numbers of samples per class.

Using Scene-LTE, our systems achieve an accuracy of 86% and a F1-score of 94.9% on the DCASE and LITIS datasets, respectively. These results surpass the best reported performance on the DCASE dataset (85% in terms of accuracy [1]) while being just marginally below the best performance on the LITIS dataset (95.6% in terms of F1-score [5]). Regarding Speech-LTE, we obtain the best accuracy, 87%, with $N = 25$ and linear kernel which is even better than that of Scene-LTE on the DCASE 2013 dataset. For the LITIS dataset, although Speech-LTE maintains good F1-score near 90% in most of the cases, these results are not as good as with those of Scene-LTE. It is also noticeable that adding more selected phone triplet categories per scene class seems to bring up the performance, however, the gains are insignificant in most of the cases. Lastly, for both DCASE and LITIS datasets, the performance of the linear systems are comparable with the nonlinear ones with $\chi^2$, hist., and RBF kernels. This is good since the linear systems are computationally much cheaper to train and evaluate compared to the nonlinear ones.

For the fusion of Scene-LTE and Speech-LTE into Fusion-LTE with the extended Gaussian-$\chi^2$ kernel, we obtain average accuracy gains of 0.8% and 2.2% compared to individual Scene-LTE and Speech-LTE with the $\chi^2$ kernel on the DCASE dataset. Similarly, the average F1-score gains on the LITIS dataset are 1.3% and 7.9%. These results indicate that speech signals are not only able to represent well the scene audio signals but also provide a valuable external source to enhance the performance of a classification system built on the scene data itself.

Finally, we present a comprehensive performance comparison of our systems and other reported results on the DCASE and LITIS datasets in Tables 4 and 5, respectively. We mark in bold where the performance of our systems outperforms all the opponents. Since the results on the LITIS dataset were reported with different metrics, i.e. average class-wise precision [23, 19], average class-wise F1-score [4, 5], and overall accuracy [4, 28], we provide our performance on all of these metrics to make a proper comparison. We also would like to notice that although there exists other works on the DCASE dataset after the challenge, we only mention here those with performance equivalent or higher than that of the best submission in the challenge. For the details of the competitive systems, please refer to the respective references. As can be seen for the DCASE dataset, our systems consistently outperform the best submission to the DCASE 2013 challenge (i.e. RNH [24]) with a large margin of about 10% and also outrun the best reported performance in [1] from 1% to 2%. For the LITIS dataset, our systems with Scene-LTE alone show better performance than most of the compared systems. Moreover, our Fusion-LTE systems set state-of-the-art performance on all evaluation metrics and outperform the best reported results by 3.7%, 0.6%, and 0.4% in terms of precision, F1-score, and accuracy, respectively.

Table 4: Performance comparison on the DCASE 2013 dataset.

| Systems | Accuracy |
|---|---|
| *Scene-LTE* | **86.0** |
| *Speech-LTE* | **87.0** |
| *Fusion-LTE* | **87.0** |
| RNH [24] | 76.0 |
| MV [24] | 77.0 |
| Human [2] | 75.0 |
| HOG [23] | 76.0 |
| AMS+LDA [1] | 85.0 |

Table 5: Performance comparison on the LITIS Rouen dataset.

| Systems | Prec. | F1-score | Acc. |
|---|---|---|---|
| *Scene-LTE* | **94.6** | 94.9 | 94.9 |
| *Speech-LTE* | 89.7 | 90.1 | 90.3 |
| *Fusion-LTE* | **95.9** | **96.2** | **96.4** |
| HOG [23] | 91.7 | – | – |
| HOG+SPD [4] | 93.3 | 92.8 | 93.4 |
| DNN+MFCC [19] | 92.2 | – | – |
| Sparse NMF [5] | – | 94.1 | – |
| Convolutive NMF [5] | – | 94.5 | – |
| Kernel PCA [5] | – | 95.6 | – |
| HOG+ProbSVM [28] | – | – | 96.0 |

### 4. CONCLUSIONS

In this paper, we present efficient schemes for acoustic scene classification. We explore the structure of the class labels by automatically learning class-label hierarchies and then the label-tree embeddings to map scene instances into the semantic space underlying the class hierarchy. We study both the label tree embedding intrinsically learned from the scene data as well as the one learned from the external TIMIT speech data. Both of them demonstrate good empirical performance on the experimental datasets, including the DCASE 2013 and LITIS Rouen datasets. Furthermore, fusing them with a simple scheme leads to state-of-the-art performance.

### 5. ACKNOWLEDGMENTS

This work was supported by the Graduate School for Computing in Medicine and Life Sciences funded by Germany's Excellence Initiative [DFG GSC 235/1].

# 6. REFERENCES

[1] S. Ağcaer, A. Schlesinger, F.-M. Hoffmann, and R. Martin. Optimization of amplitude modulation features for low-resource acoustic scene classification. In *Proc. European Signal Processing Conference (EUSIPCO)*, pages 2556–2560, 2015.

[2] D. Barchiesi, D. Giannoulis, D. Stowell, and M. Plumbley. Acoustic scene classification: Classifying environments from the sounds they produce. *IEEE Signal Processing Magazine*, 32(3):16–34, 2015.

[3] S. Bengio, J. Weston, and D. Grangier. Label embedding trees for large multi-class tasks. In *Proc. Advances in Neural Information Processing Systems (NIPS)*, pages 163–171, 2010.

[4] V. Bisot, S. Essid, and G. Richard. HOG and subband power distribution image features for acoustic scene classification. In *Proc. European Signal Processing Conference (EUSIPCO)*, pages 719–723, 2015.

[5] V. Bisot, R. Serizel, S. Essid, and G. Richard. Acoustic scene classification with matrix factorization for unsupervised feature learning. In *Proc. IEEE International Conference on Acoustics, Speech and Signal Processing (ICASSP)*, pages 6445–6449, 2016.

[6] L. Breiman. Random forest. *Machine Learning*, 45:5–32, 2001.

[7] R. Cai, L. Lu, and A. Hanjalic. Co-clustering for auditory scene categorization. *IEEE Trans. Multimedia*, 10(4):596–606, 2008.

[8] S. Chu, S. Narayanan, C.-C. J. Kuo, and M. J. Mataric. Where am I? Scene recognition for mobile robots using audio features. In *Proc. IEEE International Conference on Multimedia and Expo (ICME)*, pages 885–888, 2006.

[9] S. Deng, J. Han, C. Zhang, T. Zheng, and G. Zheng. Robust minimum statistics project coefficients feature for acoustic environment recognition. In *Proc. IEEE International Conference on Acoustics, Speech and Signal Processing (ICASSP)*, pages 8232–8236, 2014.

[10] D. P. W. Ellis. Gammatone-like spectrograms, 2009.

[11] A. J. Eronen, V. T. Peltonen, J. T. Tuomi, A. P. Klapuri, S. Fagerlund, T. Sorsa, G. Lorho, and J. Huopaniemi. Audio-based context recognition. *IEEE Trans. Audio, Speech, and Language Processing*, 14(1):321–329, 2006.

[12] W. Fisher, G. Doddington, and K. Goudie-Marshall. The DARPA speech recognition research database: Specifications and status. In *Proc. DARPA Workshop on Speech Recognition*, pages 93–99, 1986.

[13] T. Heittola, A. Mesaros, A. Eronen, and T. Virtanen. Context-dependent sound event detection. *EURASIP Journal on Audio, Speech, and Music Processing*, 2013.

[14] T. Heittola, A. Mesaros, A. J. Eronen, and T. Virtanen. Audio context recognition using audio event histogram. In *Proc. European Signal Processing Conference (EUSIPCO)*, pages 1272–1276, 2010.

[15] I. Laptev, M. Marszałek, C. Schmid, and B. Rozenfeld. Learning realistic human actions from movies. In *Proc CVPR*, pages 1–8, 2008.

[16] R. F. Lyon. Machine hearing: An emerging field. *IEEE Signal Processing Magazine*, 27(5):131–139, 2010.

[17] R. Mogi and H. Kasaii. Noise-robust environmental sound classification method based on combination of ICA and MP features. *Artificial Intelligence Research*, 2(1):107–121, 2013.

[18] A. Y. Ng, M. I. Jordan, and Y. Weiss. On spectral clustering: Analysis and an algorithm. In *Proc. NIPS*, pages 849–856, 2001.

[19] Y. Petetin, C. Laroche, and A. Mayoue. Deep neural networks for audio scene recognition. In *Proc. European Signal Processing Conference (EUSIPCO)*, pages 125–129, 2015.

[20] H. Phan, L. Hertel, M. Maass, R. Mazur, and A. Mertins. Representing nonspeech audio signals through speech classification models. In *Proc. Interspeech*, pages 3441–3445, 2015.

[21] H. Phan, L. Hertel, M. Maass, R. Mazur, and A. Mertins. Learning representations for nonspeech audio events through their similarities to speech patterns. *IEEE/ACM Trans. Audio, Speech, and Language Processing*, 24(4):807–822, April 2016.

[22] R. Radhakrishnan, A. Divakaran, and P. Smaragdis. Audio analysis for surveillance applications. In *Proc. IEEE Workshop on Applications of Signal Processing to Audio and Acoustics (WASPAA)*, pages 158–161, 2005.

[23] A. Rakotomamonjy and G. Gasso. Histogram of gradients of time-frequency representations for audio scene classification. *IEEE/ACM Trans. Audio, Speech, and Language Processing*, 23(1):142–153, 2015.

[24] D. Stowell, D. Giannoulis, E. Benetos, M. Lagrange, and M. D. Plumbley. Detection and classification of acoustic scenes and events. *IEEE Trans. Multimedia*, 17(10):1733–1746, 2015.

[25] X. Valero and F. Alías. Gammatone cepstral coefficients: biologically inspired features fro non-speech audio classification. *IEEE Trans. Multimedia*, 17(6):1684–1689, 2012.

[26] D. Wang and G. J. Brown. *Computational Auditory Scene Analysis: Principles, Algorithms, and Applications*. Wiley-IEEE Press, 2006.

[27] Y. Xu, W. J. Li, and K. K. Lee. *Intelligent Wearable Interfaces*. Hoboken, NJ: Wiley, 2008.

[28] J. Ye, T. Kobayashi, M. Murakawa, and T. Higuchi. Acoustic scene classification based on sound textures and events. In *Proc. ACM Multimedia*, pages 1291–1294, 2015.